\documentclass[twocolumn,showpacs,preprintnumbers,amsmath,amssymb]{revtex4}
\usepackage{graphicx}% Include figure files
\usepackage{dcolumn}% Align table columns on decimal point
\usepackage{bm}% bold math

\begin{document}

%\preprint{Chiba/Physics}

\title{Constraints on Neutrino-Nucleon Interactions at energies of 1 EeV 
with the IceCube Neutrino Observatory}
% Force line breaks with \\

\author{Shigeru Yoshida}
\email{syoshida@hepburn.s.chiba-u.ac.jp}
\homepage{http://www.ppl.phys.chiba-u.jp/~syoshida}
\affiliation{%
Department of Physics, Graduate School of Science, Chiba University Chiba 263-8522, Japan
}%

\date{\today}% It is always \today, today,

\begin{abstract}
A search for extremely high energy cosmic neutrinos 
has been carried out with the IceCube Neutrino Observatory. 
The main signals in the search are neutrino-induced
energetic charged leptons and their rate depends on the neutrino-nucleon cross section.
The upper-limit on the neutrino flux has implications for possible new physics beyond the 
standard model such as the extra space-time dimension scenarios which
lead to a cross section much higher than the standard particle physics 
prediction. In this study we constrain the neutrino-nucleon cross section at energies beyond $10^9$ GeV
with the IceCube observation. 
The constraints are obtained as a function of the extraterrestrial neutrino 
flux in the relevant energy range, which accounts for the astrophysical 
uncertainty of neutrino production models.
\end{abstract}

\maketitle

%=====================================================
%=====================================================
\section{\label{sec:intro} Introduction}
%=====================================================
%=====================================================

High energy cosmic neutrino observations provide a rare opportunity
to explore the neutrino-nucleon ($\nu$N) interaction behavior 
beyond energies accessible by the present accelerators. 
These neutrinos interact during their propagation in the earth and 
produce energetic muons and taus. These secondary leptons reach 
underground neutrino detectors and leave detectable signals. 
The detection rate is, therefore,
sensitive to neutrino-nucleon interaction probability. 
The center-of-mass energy of the collision, $\sqrt{s}$, 
is well above $\sim 10$ TeV for cosmic neutrino energies 
on the order of 1 EeV ($=10^9$ GeV).
This is a representative energy range for the bulk of the GZK cosmogenic neutrinos, 
generated by the interactions between the highest energy cosmic ray nucleons
and the cosmic microwave background photons~\cite{berezinsky69}. 

The $\nu$N collision cross section can vary greatly if 
non-standard particle physics beyond the Standard Model (SM)
is considered in the high energy regime of $\sqrt{s}\gg$ TeV.
The extra-dimension scenarios, for example, have predicted
such effects~\cite{jain00,tyler01}. In these scenarios, 
the virtual exchange of Kaluza-Klein graviton~\cite{jain00},
or microscopic black hole production~\cite{Anchordoqui02} 
leads to a substantial increase of the neutrino-nucleon cross section 
by more than two orders of magnitude
above the SM prediction. 
The effect would be sizable enough to affect the expected
annual event rate ($O(0.1-1)$) of the GZK neutrinos 
in the $\sim$ km$^3$ instrumentation volume of an underground neutrino telescope 
such as the IceCube observatory. Thereby 
the search for extremely-high energy (EHE) cosmic neutrinos
leads to constraints on non-standard particle physics~\cite{BH02}.

The IceCube neutrino observatory has already begun 
EHE neutrino hunting with the partially deployed 
underground optical sensor array~\cite{IceCube}. The 2007 partial 
IceCube detector realized a $\sim 0.7$ km$^2$
effective area for muons with $10^9$ GeV
and recently placed a limit on the flux of EHE neutrinos
approximately an order of magnitude higher
than the expected GZK cosmogenic neutrino intensities
with 242 days of observation~\cite{IC22-EHE}. Since new particle physics
may vary the cross section by more than an order of magnitude
as we noted above, this result should already imply a meaningful
bound on the $\nu$N cross section. In this paper,
we study the constraint on the $\nu$N cross section 
($\sigma_{\nu\rm{N}}$) 
by the null detection of EHE neutrinos with the 2007
IceCube observation.
A model-independent bound is derived by estimating 
the lepton intensity at the IceCube depth
with the SM cross section scaled by a constant. 
The constraint is displayed in the form of the excluded region 
on the plane of the cosmic neutrino flux and $\sigma_{\nu{\rm N}}$. 
It is equivalent to an upper-bound on $\sigma_{\nu{\rm N}}$
for a given flux of astrophysical EHE neutrinos.
We also study the model-dependent constraint on the microscopic black hole
creation by neutrino-nucleon collision predicted in 
the extra-dimension scenario~\cite{BH02}.
We calculate the fluxes of leptons propagating in the earth
including the black hole cross section and the final states
to estimate expected event rate in an equivalent IceCube 2007 measurement
as a function of extraterrestrial neutrino intensity.
The null detection of signal candidates leads to a 
constraint on this particular scenario.

There are several works on model-independent 
upper bounds of $\sigma_{\nu\rm{N}}$ using the observational limit 
of EHE neutrino flux in the literature. 
Refs.~\cite{tyler01,Anchordoqui02, Anchordoqui05} derived the bound using
the results of horizontal air shower search by AGASA~\cite{yoshida01}
and Fly's Eye~\cite{Baltrusaitis85}. Refs.~\cite{Anchordoqui05,Barger06}
set the limit based upon the flux bound by the RICE experiment~\cite{rice}.
Our approach in the present study is different mainly in two respects.
The previous works assumed the GZK cosmogenic neutrino bulk as
the {\it guaranteed} beam and deduced the cross section limit
using the GZK neutrino intensity. Here extraterrestrial neutrino intensity
is considered as a free parameter. This method is an application
of the technique to derive the 
flux limit based upon the quasi-differential event rate~\cite{IC22-EHE,rice,auger}, 
which is independent of specific neutrino flux models.
As EHE cosmic ray composition and their origin are still quite uncertain, 
this approach provides more appropriate conservative 
limits on  $\sigma_{\nu\rm{N}}$.
It also allows estimation of the minimum intensity of neutrino flux required
to constrain the cross section. Another difference is that
the previous works introduced the simplification that 
event rate solely depends on rates of electromagnetic
or hadronic cascades directly initiated by neutrinos inside the effective
volume of the detector. This is in fact
a good approximation for the RICE experiment which is sensitive to radio emission
from shower events. However, underground neutrino telescopes such as IceCube
have larger effective area for through-going muons and taus in EHE neutrino
search~\cite{halzen01,reno04,yoshida04}. 
This study of the model-independent
limit includes calculation of not just intensities of neutrinos but also the
secondary muon and tau fluxes reaching the detection volume
for a given $\sigma_{\nu\rm{N}}$
and includes their contributions in the overall event rate.

The paper is outlined as follows: First we discuss
the model-independent constraint in Sec.~\ref{sec:model-indep-bound}. 
The method to calculate the neutrino and 
the secondary lepton propagation
from the earth's surface to the IceCube detector depth
is described. The fluxes for different strengths
of $\sigma_{\nu\rm{N}}$ are calculated
and the resultant constraint
is shown for both $\sigma_{\nu\rm{N}}$ and the cosmic neutrino
flux at neutrino energies of 1 and 10 EeV, respectively.
Sec.~\ref{sec:bh-bound} describes the constraint
on the microscopic black hole production by neutrino-nucleon
interaction as an example of the model dependent bound
on $\sigma_{\nu\rm{N}}$. Fluxes of muons and taus from evaporation
of black holes produced in the neutrino-nucleon collision
in the earth are calculated. Their contributions, as well as 
those from contained hadronic showers induced directly
by the evaporation, would give an observable event rate
in the IceCube 2007 measurement, and thereby put constraints
on the black hole scenario.
We summarize our conclusions in Sec.~\ref{sec:summary}.

%=====================================================
%=====================================================
\section{\label{sec:model-indep-bound} Model Independent Constraint 
on the neutrino-nucleon cross section}
%=====================================================
%=====================================================

The flux limit obtained by the present IceCube observations
allows us to place an upper bound on the neutrino-nucleon cross section
in a model independent manner; new physics cannot increase $\sigma_{\nu\rm{N}}$
too much, otherwise EHE neutrinos would have produced observable events.
As an underground neutrino telescope is sensitive to not just shower events
induced from neutrinos, but to through-going muons and taus generated by
the neutrino-nucleon scattering, one must understand how much
fluxes of these leptons reaching an underground detection volume 
is increased with $\sigma_{\nu\rm{N}}$. In this section,
we first discuss our method to calculate 
intensities of neutrinos, muons, and taus
at the underground depth of the IceCube observatory
for a wide range of $\sigma_{\nu\rm{N}}$ strength, followed by a
description of how they would contribute to the event rate.
Finally the constraint on both $\sigma_{\nu\rm{N}}$ and
cosmic neutrino flux is described together with the relevant discussions.

%=====================================================
%=====================================================
\subsection{\label{subsec:indep-method} The Method}
%=====================================================
%=====================================================

Given a neutrino flux at the surface of the Earth, the neutrino and charged lepton fluxes at the IceCube depth 
are calculated by the coupled transportation equations~\cite{yoshida04}:
%=====================================================
{\small
\begin{eqnarray}
{dJ_{\nu}\over dX} &=& -N_A\sigma_{\nu N, CC+NC}J_{\nu}+\
{m_l\over c\rho\tau^d_l}\int dE_l {1\over E_l} 
{dn^d_l\over dE_{\nu}}J_{l}(E_l)\nonumber\\
&& + N_A \int dE^{'}_{\nu} \
{d\sigma_{\nu N, NC}\over dE_{\nu}}J_{\nu}(E^{'}_{\nu})\nonumber\\
&& + N_A \int dE^{'}_{l} \
{d\sigma_{l N, CC}\over dE_{\nu}}J_l(E^{'}_{l})\label{eq:transport1}\\
{dJ_{l}\over dX} &=& -N_A\sigma_{l N}J_{l}-\
{m_l\over c\rho\tau^d_l E_l}J_{l}\nonumber\\
&&+ N_A \int dE^{'}_{\nu} {d\sigma_{\nu N, CC}\over dE_{l}}\
J_{\nu}(E^{'}_{\nu})\nonumber\\
&& + N_A \int dE^{'}_{l} {d\sigma_{l N}\over dE_{l}}J_l(E^{'}_{l})\nonumber\\
&&+{m_l\over c\rho\tau^d_l}\int dE^{'}_l \
{1\over E^{'}_l} {dn^d_l\over dE_l}J_{l}(E^{'}_{l}),
\label{eq:transport2}
\end{eqnarray}
}
%================================================
where $J_{l}=dN_l/dE_l$ and $J_{\nu}=dN_{\nu}/dE_{\nu}$ are
differential
fluxes of charged leptons (muons and taus) and neutrinos, respectively.
$X$ is the column density, 
$N_A$ is the Avogadro's number,
$\rho$ is the local density of the medium (rock/ice) 
in the propagation path, 
$\sigma$ is the relevant interaction cross section,
$dn^d_l/dE$
is the energy distribution of the decay products which is derived from
the decay rate per unit energy,
$c$ is the speed of light,
$m_l$ and $\tau^d_l$ are the mass and the decay life time 
of the lepton $l$, respectively. 
CC(NC) denotes the charged (neutral) current interaction.
In this study we scale $\sigma_{\nu {\rm N}}$ to
that of the SM prediction with the factor N$_{\rm scale}$, {\it i.e.},
$\sigma_{\nu {\rm N}} \equiv {\rm N_{scale}}\sigma_{\nu {\rm N}}^{{\rm SM}}$.
It is an extremely intensive computational task to resolve
the coupled questions above for every possible value of 
$\sigma_{\nu {\rm N}}$. To avoid this difficulty,
we introduce two assumptions to decouple calculation of 
$J_{\nu}$ from the charged lepton transportation equation.
The first is that distortion of the neutrino spectrum 
by the neutral current reaction is small and 
the other is that regeneration of neutrinos due to muon and tau decay
and their weak interactions is negligible. 
These are very good approximations in the energy region
above $10^8$ GeV where even tau is unlikely to decay before reaching the IceCube instrumentation
volume. Then the neutrino flux is simply given by
the beam dumping factor as
%=====================================================
\begin{equation}
J_{\nu}({\rm E_\nu, X_{IC}}) = 
J_{\nu}({\rm E_\nu, 0}) 
e^{-N_{{\rm scale}}\sigma_{\nu {\rm N}}^{SM,CC}{\rm X_{IC}}},
\label{eq:nu-transport}
\end{equation}
%=====================================================
where ${\rm X_{IC}}$ is column density of the propagation path
from the earth surface to the IceCube depth.
The charged lepton fluxes, $J_{l=\mu,\tau}(\rm{E_l,X_{IC}})$, 
are obtained as
%=====================================================
{\small
\begin{equation*}
J_{\mu,\tau}({\rm E_{\mu,\tau},X_{IC}}) = N_A\int\limits_0^{{\rm X_{IC}}}\
dX \int dE^{'}_{\mu,\tau} {dN_{\mu,\tau}\over dE_{\mu,\tau}}\
({\rm E_{\mu,\tau}^{'}\to E_{\mu,\tau}})
\end{equation*}
\begin{equation}
\int d{\rm E_\nu}{\rm N_{scale}}{d\sigma_{\nu N}^{SM, CC}\over dE_{\mu,\tau}^{'}}\
 J_{\nu}({\rm E_\nu, 0})e^{-N_{{\rm scale}}\sigma_{\nu {\rm N}}^{SM, CC}{\rm X}}.
\label{eq:lepton-transport}
\end{equation}
}
%=====================================================
Here $dN_{\mu,\tau}/dE_{\mu,\tau}({\rm E_{\mu,\tau}^{'}\to E_{\mu,\tau}})$ 
represents distributions of muons and taus with energy of 
${\rm E_{\mu,\tau}}$ at ${\rm X_{IC}}$
created by $\nu{\rm N}$ collisions at depth ${\rm X}$ with an energy ${\rm E_{\mu,\tau}^{'}}$.
This is calculated in the transportation equation, Eq.~\ref{eq:transport2},
with a replacement of $J_{\nu}(E^{'}_{\nu})$ by Eq.~\ref{eq:nu-transport}.

%================================================
\begin{figure}[!t]
 \centering
 \includegraphics[width=.4\textwidth,clip=true]{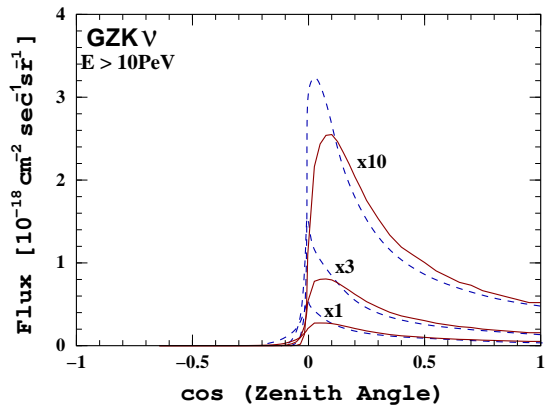}
 \caption{Integral fluxes of the muon and taus above 10 PeV ($=10^7$ GeV) 
 at IceCube depth ($\sim$ 1450m) 
 for GZK cosmogenic neutrinos~\cite{yoshida97}.
 The solid lines represent muons while the dashed lines represent taus. 
 Numbers on each of the curves are the multiplication factors (N$_{\rm scale}$) that enhance 
 the standard $\nu$N cross section~\cite{gandhi96} 
 in the relevant calculations.}
 \label{fig:AngularFlux}
\end{figure}
%================================================

Calculation of the neutrino and the charged lepton fluxes with this method
is feasible for a wide range of ${\rm N_{scale}}$  
without any intensive computation. 
A comparison of the calculated fluxes with those obtained without
the introduced simplification for a limited range of ${\rm N_{scale}}$
indicates that the relative difference
we found in the resultant $J_{\nu,\mu,\tau}(X_{IC})$ is within 40\%.
Since this analysis involves an {\it order} 
of magnitude of increase in $\sigma_{\nu{\rm N}}$, 
the introduced approximations provide sufficient accuracy 
for the present study.

FIG.~\ref{fig:AngularFlux} shows the calculated intensities of the secondary
muons and taus for various  ${\rm N_{scale}}$ factors. 
Here the primary neutrino spectrum is assumed to follow the GZK cosmogenic spectrum
and flux calculated in Ref.~\cite{yoshida97} assuming an all-proton
cosmic-ray composition with a moderately strong source evolution,
$(1+z)^m$ with $m=4$ extending to $z=4$.
One can see that the intensity is nearly proportional to
${\rm N_{scale}}$ as expected since the interaction probability to
generate muons and taus linearly depends on $\sigma_{\nu{\rm N}}$.
It should be pointed out, however, that the dependence starts 
to deviate from  the complete linearity 
when the propagation distance is comparable 
to the mean free path of neutrinos,
as one can find in the case of ${\rm N_{scale}}=10$ in the figure. 
This is because the neutrino beam dumping factor 
in Eq.~\ref{eq:nu-transport} becomes
significant under this circumstances. 

The flux yield of leptons at the IceCube depth, $Y_\nu^l$ 
($l=\nu^{'}$s, $\mu,\tau$), originating from
neutrinos with the same energy at the earth's surface, $E_\nu^s$,
is given by Eq.~\ref{eq:lepton-transport} for muons and taus,
by Eq.~\ref{eq:nu-transport} for neutrinos, with an insertion of
$J_{\nu}(E_\nu,0)=\delta(E_{\nu}-E_{\nu}^s)$. The resultant event rate
per neutrino energy decade is then obtained by~\cite{IC22-EHE,rice,auger},
%=====================================================
{\small
\begin{equation*}
N_\nu(E_\nu^s) =\sum\limits_{\nu=\nu_e,\nu_\mu,\nu_\tau}{1\over 3}\
{dJ_{\nu_e+\nu_\mu+\nu_\tau}\over d\log E_\nu}\left(E_\nu^s\right)\int d\Omega
\end{equation*}
\begin{equation}
\sum\limits_{l=\nu_e,\nu_\mu,\nu_\tau,\mu,\tau}\int dE_l A_l(E_l) \
Y_\nu^l(E_\nu^s,E_l,{\rm X_{IC}(\Omega), N_{scale}})
\label{eq:event_rate}
\end{equation}
}
%{\bf I think there shouldn't be a sum in the top row. Is this what you meant?}
%\begin{equation*}
%N_\nu(E_\nu^s) ={1\over 3}\
%{dJ_{\nu_e+\nu_\mu+\nu_\tau}\over d\log E_\nu}\left(E_\nu^s\right)\int d\Omega
%\end{equation*}
%{\bf Or maybe this? Otherwise it seems like you're adding the flux from all three flavors three times}
%\begin{equation*}
%N_\nu(E_\nu^s) =\sum\limits_{\nu=\nu_e,\nu_\mu,\nu_\tau}{1\over 3}\
%{dJ_{\nu}\over d\log E_\nu}\left(E_\nu^s\right)\int d\Omega
%\end{equation*}
%=====================================================
where $A_l$ is the effective area of the IceCube to detect the lepton $l$.
In this equation above, the $\l=\mu,\tau$ terms represent
the through-going track events while contribution of events
directly induced by neutrinos inside the detection volume
is represented by the terms $\l=\nu_e,\nu_\mu,\nu_\tau$.
The effective area for $\nu^{'}$s, $A_{\nu}$, is proportional to
$\sigma_{\nu {\rm N}}$ {\it i.e.}, ${\rm N_{scale}}$
so the rate of contained shower events is linearly dependent
on the neutrino-nucleon scattering probability.
Note that the differential limit of the neutrino flux is 
given by Eq.~\ref{eq:event_rate}
for ${\rm N_{scale}}=1$ with $N_\nu={\bar \mu_{90}}$ 
which corresponds to the 90 \%
confidence level average upper limit. This calculation is valid when
the cosmic neutrino flux $J_\nu$ and the cross section $\sigma_{\nu {\rm N}}$
do not rapidly change over a decade of neutrino energy around $E_\nu^s$.
Limiting $\sigma_{\nu {\rm N}}$ in the present analysis corresponds to
an extraction of  the relation between ${\rm N_{scale}}$ and 
the (unknown) cosmic neutrino flux $J_{\nu_e+\nu_\mu+\nu_\tau}$
yielding $N_\nu={\bar \mu_{90}}$. The obtained constraints on 
$\sigma_{\nu {\rm N}}$
is represented as a function of $J_{\nu_e+\nu_\mu+\nu_\tau}$ for a given
energy of $E_{\nu}^s$. 
It consequently accounts for astrophysical uncertainties
on the cosmic neutrino flux.

In scenarios with extra dimensions and strong gravity, Kaluza-Klein gravitons
can change only the neutral current (NC) cross section because
gravitons are electrically neutral. Any scenarios belonging to this category
can be investigated by scaling only $\sigma_{\nu {\rm N}}^{NC}$ in the present
analysis. The event rate calculation by Eq.~\ref{eq:event_rate} 
is then performed
for $Y_\nu^l({\rm N_{scale}=1})$ with effective area for $\nu$'s, $A_{\nu}$,
enhanced by
$(\sigma_{\nu {\rm N}}^{SM,CC}+{\rm N_{scale}}\sigma_{\nu {\rm N}}^{SM,NC})/
(\sigma_{\nu {\rm N}}^{SM,CC}+\sigma_{\nu {\rm N}}^{SM,NC})$
since the rate of detectable events via the NC reaction by IceCube
is proportional to $\sigma_{\nu {\rm N}}^{NC}$. We also show the constraint
in this case.

%=====================================================
%=====================================================
\subsection{\label{sec:indep-reults} Results}
%=====================================================
%=====================================================

%================================================
\begin{figure}[tb]
 \centering
 \includegraphics[width=.4\textwidth,clip=true]{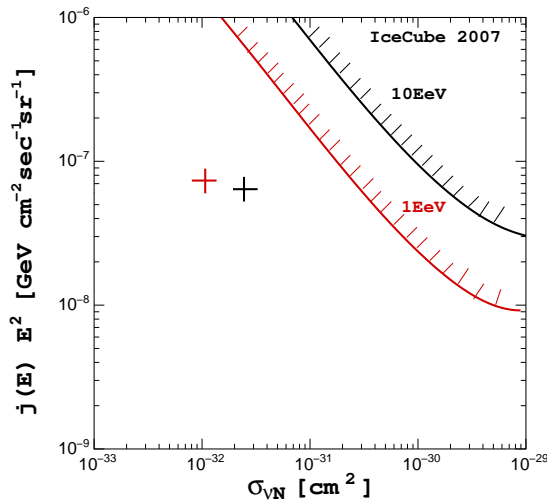}
 \caption{Constraints on the all-flavor sum of  cosmic neutrino flux and 
 the charged current $\nu$N cross section
 based on the null detection of neutrino signals
 by the IceCube 2007 observation. 
 The right upper region is excluded
 by the present analysis. The cross points provide reference points
 where the standard cross section~\cite{gandhi96} and the expected GZK cosmogenic 
 neutrino fluxes~\cite{kalashev02} is located.}
 \label{fig:Constraint-CC}
\end{figure}
%================================================
In this analysis we use the IceCube observation results with 242 days data in 2007
to limit $\sigma_{\nu {\rm N}}$ using Eq.~\ref{eq:event_rate}. No detection
of signal candidates in the measurement has led to an upper limit
of the neutrino flux of $1.4\times 10^{-6}$ GeV cm$^{-2}$ sec$^{-1}$ sr$^{-1}$~\cite{IC22-EHE}
in the energy range from $3\times 10^7$ to $3\times 10^9$ GeV.
The effective area $A_l$ is $\sim$0.7 km$^2$ for $\mu$, $\sim$
0.4 km$^2$ for $\tau$, and $3\times 10^{-4}$ km$^2$ for $\nu^{'}$s~\cite{IC22-EHE}.
Constraints on $\sigma_{\nu {\rm N}}$ are then derived 
with Eq.~\ref{eq:event_rate}.
The results for $E_\nu^s=10^9$ and $10^{10}$ GeV are shown
in FIG.~\ref{fig:Constraint-CC}. Enhancing the charged current cross section
by more than a factor of 30 for $E_\nu= 1$ EeV ($10^{9}$ GeV) is disfavored 
if the astrophysical neutrino intensities are around $\sim 10^{-7}$ 
GeV cm$^{-2}$ sec$^{-1}$ sr$^{-1}$, 
near the upper bound of the GZK cosmogenic neutrino bulk.
%the expected range of
%the GZK cosmogenic neutrino bulk. 
Note that neutrino-nucleon collision
with $E_\nu= 1$ EeV corresponds to $\sqrt{s}\sim 40$ TeV and the present limit on 
$\sigma_{\nu {\rm N}}$ would place a rather strong constraint 
on scenarios with extra dimensions and strong gravity,
although more accurate estimation requires studies with a model-dependent approach 
which implements the cross section and the final-state particles from the collision
predicted by a given particle physics model.
Taking into account uncertainty on the astrophysical neutrino fluxes,
any model that increases the neutrino-nucleon cross section 
to produce charged leptons 
by more than two orders of magnitude at $\sqrt{s}\sim 40$ TeV is
disfavored by the IceCube observation. However, we should point out that
the IceCube 2007 data could not constrain the charged current cross section 
if the intensity of cosmic neutrinos in the relevant energy region is fewer than
$\sim 10^{-8}$ GeV cm$^{-2}$ sec$^{-1}$ sr$^{-1}$, within the lower range of
prediction for the cosmogenic neutrino fluxes~\cite{ahlers2010}. 
Absorption effects in the earth becomes sizable
in this case, resulting in less sensitivity to the cross section.
This limitation will be improved for larger detection area of the full IceCube detector.

%================================================
\begin{figure}[tb]
 \centering
 \includegraphics[width=.4\textwidth,clip=true]{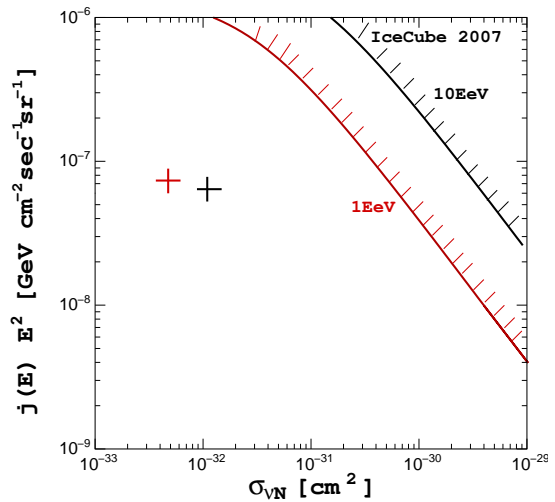}
 \caption{Constraints on the all-flavor sum of cosmic neutrino flux and 
 the neutral current $\nu$N cross section for the scenario
 that only the neutral current reaction is enhanced by a new physics
 beyond the standard model.
 The right upper region is excluded
 by the present analysis. The crosses provide reference points
 for the standard cross section~\cite{gandhi96} and the expected GZK cosmogenic 
 neutrino fluxes~\cite{kalashev02}.}
 \label{fig:Constraint-NC}
\end{figure}
%================================================

FIG.~\ref{fig:Constraint-NC} shows the constraints when only the NC cross section
is varied. Enhancement of $\sigma_{\nu {\rm N}}^{NC}$ by a factor beyond
100 at $\sqrt{s}\sim 40$ TeV is disfavored, but this strongly depends on
the cosmic neutrino flux one assumes. Because the NC interaction does not
absorb neutrinos during their propagation though the earth, 
the cross section could be bounded
even in the case when the neutrino flux is small, 
but the limit becomes rather weak; the allowed
maximum enhancement factor is on the order of $\sim 10^3$.

%=====================================================
%=====================================================
\section{\label{sec:bh-bound} Constraint 
on the microscopic black hole production }
%=====================================================
%=====================================================

%================================================
\begin{figure*}[t]
\begin{tabular}{cc}
 \includegraphics[width=.4\textwidth,clip=true]{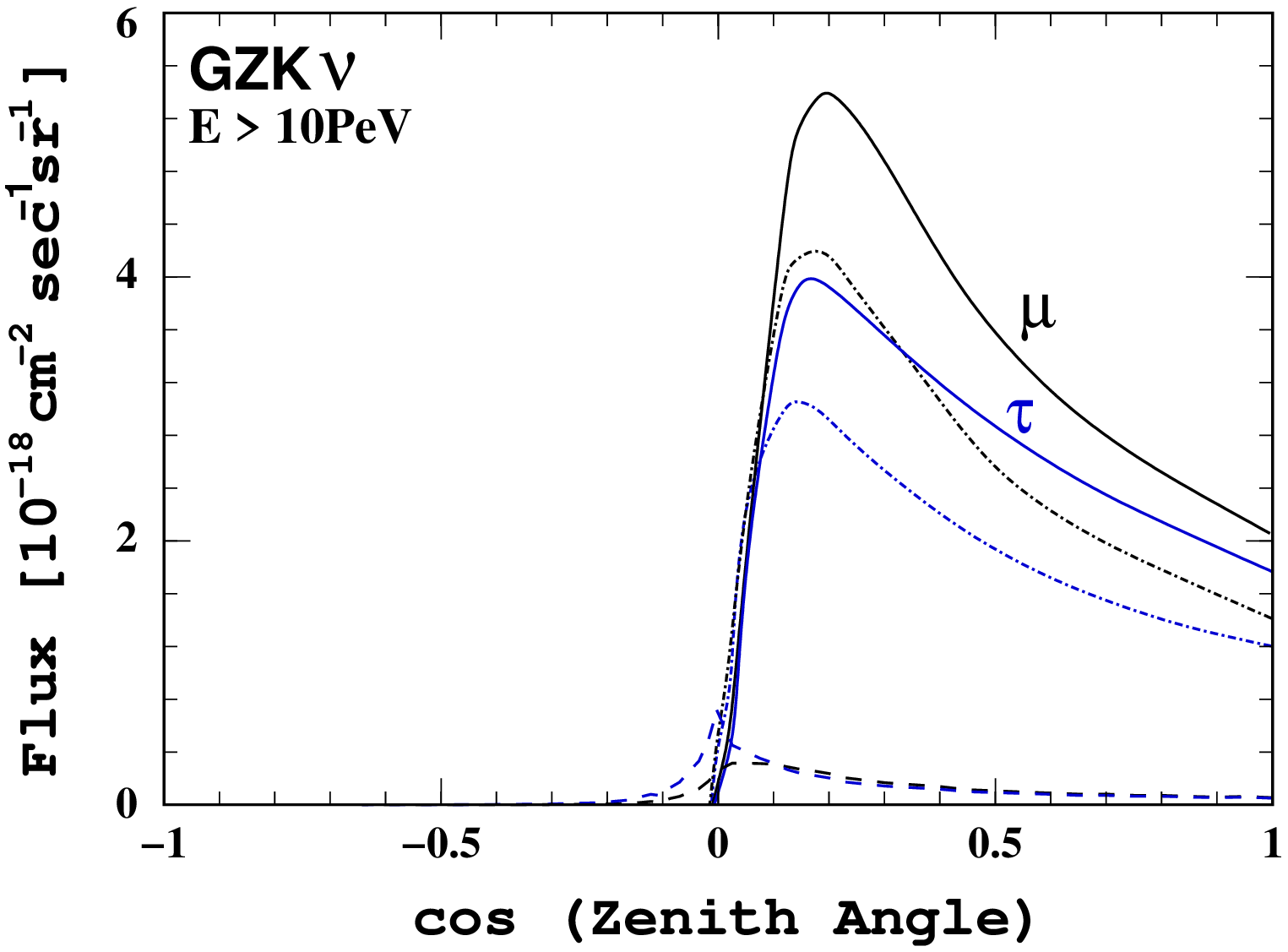} & \
 \includegraphics[width=.45\textwidth,clip=true]{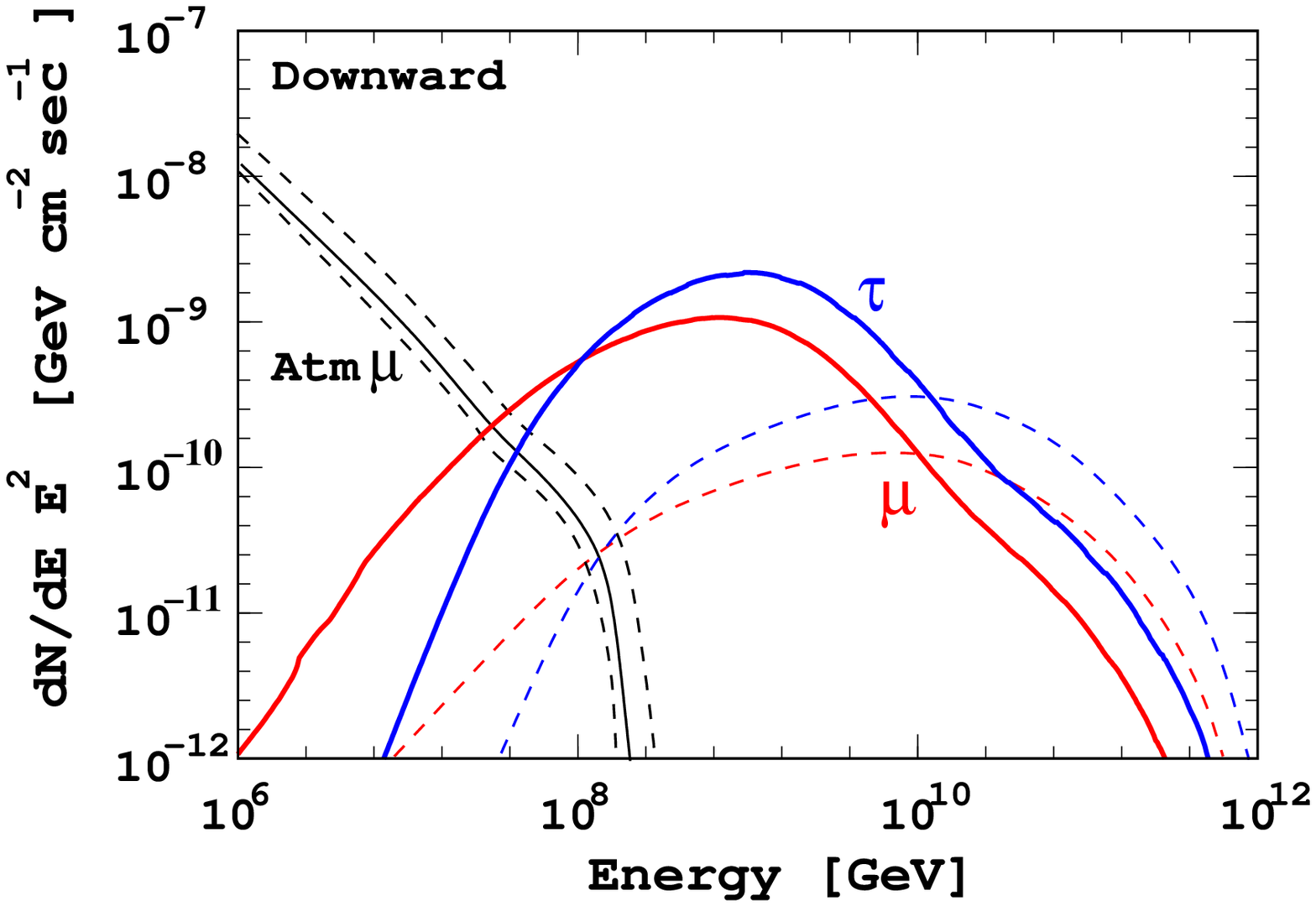}\\
\end{tabular}
 \caption{Left: Integral fluxes of the muon and taus above 10 PeV ($=10^7$ GeV) 
 at IceCube depth ($\sim$ 1450m) 
 of the GZK cosmogenic neutrinos~\cite{yoshida97} in case
 of the microscopic black hole creation scenario~\cite{BH02}
 for ($M_D$,$x_{min}$)=(1TeV,1) (solid), (1TeV,3) (dot-dash).
 The dash line corresponds to the intensities obtained by
 the SM $\nu$N cross section~\cite{gandhi96}. 
 Right: Energy spectra of the GZK $\nu$ induced muons and taus at IceCube depth 
 with downgoing ({\it i.e.} $\cos({\rm zenith})\geq 0$) geometry
 expected by the black hole model for ($M_D$,$x_{min}$)=(1TeV,1).
 The spectra produced by the SM cross section are also shown 
 as the dashed lines for comparison. 
 The curve labeled ``Atm $\mu$'' represents
 the atmospheric muon intensity estimated by the IceCube observation
 with its uncertainty expressed by the two dash lines~\cite{icrc07}.}
 \label{fig:BHFlux}
\end{figure*}
%================================================

A constraint on a specific physics model that enhances
the neutrino-nucleon cross section is obtained by
the same procedure for the model independent bound,
except the transport equations, Eqs.~\ref{eq:transport1} 
and \ref{eq:transport2},
would have total and differential neutrino cross section
provided by both SM and the new model. Here we study the model of black hole creation as a possible consequence
of low-scale gravity that may occur if space-time has more than
four dimensions. We use the predicted cross section
of lack hole production via the neutrino-nucleon scattering
described by Ref.~\cite{BH02}, parametrized by 
the Planck scale $M_D$, the ratio of the minimal black hole mass
to the Planck scale $x_{min}$, and the space-time dimension $D=4+n$.
In this paper $n=6$, and $M_D=1 \rm{TeV}$ are assumed as
representative numbers. The resultant cross section
may exceed SM interaction rates by two orders of magnitude or even
greater. Therefore, the model independent bound shown in the previous section
indicated that the 2007 IceCube observation should be already
sensitive to some of the parameter space in the black hole creation model.

The final states in the neutrino-nucleon scattering 
in this model are quite different from the SM case.
Black holes evaporate and generate multiple particles
of all kinds, like leptons, quarks, gluons, and bosons. 
These products are distributed according
to the number of degrees of freedom. Consequently, the
average number of muons and taus, $N_{\mu+\tau}$,
are $1/30$ of all particle average multiplicity ${\bar N}$,
which is also determined by the specific model.
As ${\bar N}\sim 10$ at neutrino energy of $E_{\nu}= 1$ EeV,
multiple muon or tau production would very rarely occur.
Then the {\it effective} differential cross section $d\sigma_{\nu\rm{ N}}/dE_{\mu, \tau}$
in the transport equation Eqs.~\ref{eq:transport1} and \ref{eq:transport2} 
in the black hole model is represented by

%=====================================================
%{\small
\begin{equation}
{d\sigma_{\nu\rm{ N}}\over dE_{\mu, \tau}} = {N_{\mu+\tau}(E_\nu)\over 2}
\sigma_{\nu\rm{N}}(E_\nu){{\bar N(E_\nu)}\over 2E_\nu}
\end{equation}
%}
%=====================================================
with $0\leq E_{\mu,\tau}\leq 2E_\nu/{\bar N}$. We take ${\bar N}$ from
Ref.~\cite{BH02} in the present calculation. In this specific scenario,
a muon or a tau carries a small fraction ($1/{\bar N}\sim 0.1$) 
of incoming neutrino energy $E_\nu$ in average,
in contrast to the SM collision that takes away $1-y \sim 0.8$ of 
neutrino energy by a generated charged lepton. 

Solving the transport equations gives the intensities of secondary muons
and taus, which are shown in FIG.~\ref{fig:BHFlux}.
One can find in the zenith angle distribution (the left panel)
that the intensities are increased by more than two orders
of magnitude above the SM case. The large increase of 
$\sigma_{\nu\rm{N}}$ enhances down-going event rates while
the upgoing muon and tau rates are more suppressed.
The zenith angle distribution is consistent with the original work 
in Ref.~\cite{BH02}. It should also be noted that
the energy spectra is substantially modified
from those in the SM case (the right panel).
The peak energy is around 1 EeV, an order of magnitude lower than
the SM spectrum, reflecting the fact that a smaller fraction
of neutrino energy is channeled into muons and taus via the black hole
evaporation. The peak happens to match the most sensitive energy
region in the IceCube EHE neutrino search~\cite{IC22-EHE}.

%================================================
\begin{figure}[tb]
 \centering
 \includegraphics[width=.4\textwidth,clip=true]{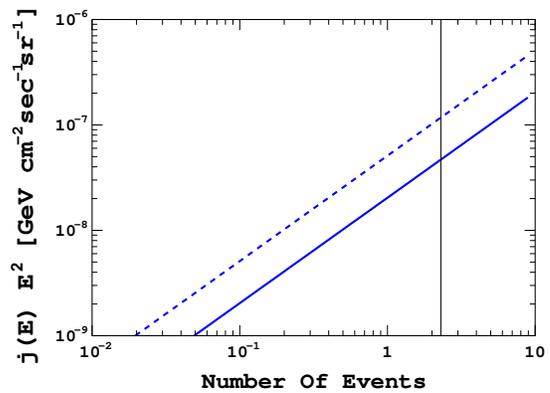}
 \caption{Number of events as a function of extraterrestrial neutrino
  flux at 1 EeV for ($M_D$,$x_{min}$)=(1TeV,1) (solid), (1TeV,3) (dash).
  The 90 \% C.L. line determined by the Poisson statistics 
  is also shown as the vertical line for reference.}
 \label{fig:FluxBoundBG}
\end{figure}
%================================================

Because $\sigma_{\nu\rm{N}}$ is solely predicted by the specific model, 
the model-dependent constraints on $\nu$N interactions 
is represented in the plane of 
extraterrestrial neutrino flux and the number of events
the IceCube 2007 run would have detected. 
FIG.~\ref{fig:FluxBoundBG} shows the number of events as a function
of the neutrino intensity at energy of 1 EeV, if the microscopic black hole evaporation
occurs as in Ref.~\cite{BH02}. The Poisson statistics then determines the upper-limit
of neutrino flux that can be still consistent with the null observation by IceCube.
It is indicated that
the neutrino intensity of $10^{-7}$ GeV cm$^{-2}$ sec$^{-1}$ sr$^{-1}$ is disfavored
in this scenario. More parameter space of $M_D$ and $x_{min}$ will be further constrained
by near future observation with IceCube whose detection volume is rapidly growing 
with increase of number of the detectors in operation.

%=====================================================
%=====================================================
\section{\label{sec:summary} Conclusions}
%=====================================================
%=====================================================

The IceCube 2007 observation indicated that any scenario to enhance
either the NC  or both the NC and CC equivalent cross section by more than
100 at $\sqrt{s}\sim 40$ TeV is unlikely 
if sum of the all three flavors of astrophysical neutrino fluxes
are greater than $\sim 3\times 10^{-8}$ GeV cm$^{-2}$ sec$^{-1}$ sr$^{-1}$ in EeV region.
Many models of the GZK cosmogenic neutrinos exist to predict this flux range,
thus the present constraints limit new particle physics beyond SM,
unless the extraterrestrial neutrino intensity is smaller than expectation.
The example of the model-dependent bound on $\sigma_{\nu{\rm N}}$
has been also shown for the microscopic black hole evaporation scenario.
A high cosmic neutrino intensity constrains the parameter space
of the black hole creation. Future observation by the rapidly growing IceCube detectors
will strongly limit particle physics models which predict an increase
of neutrino-nucleon interaction probability.

%=================================================================
%=================================================================
\begin{acknowledgments}
We wish to acknowledge the IceCube collaboration
for useful discussions and suggestions.
We thank Jonathan Feng
for helpful discussions on the physics models
beyond the standard model and 
providing numerical data of the cross section 
and multiplicity distribution predicted by
the micro black hole creation scenario.
We also wish to thank Lisa Gerhardt
for helpful comments on the manuscript.
This work was supported in part by
the Grants-in-Aid in 
Scientific Research from the JSPS (Japan Society
for the Promotion of Science) in Japan.

\end{acknowledgments}

%=================================================================
%=================================================================
%=================================================================

%=================================================================

\end{document}